\documentclass[fleqn,usenatbib]{mnras}

\usepackage{graphicx}	
\usepackage{amsmath}	
\usepackage{multicol}        
\usepackage{bm}		
\usepackage{pdflscape}	

\usepackage{amsmath}
\usepackage{amssymb,microtype,siunitx,booktabs}
\usepackage{gensymb}
\usepackage{cleveref}
\usepackage{multicol}
\usepackage{graphicx}

\usepackage{lineno}

\usepackage[T1]{fontenc}
\usepackage{ae,aecompl}

\usepackage{newtxtext,newtxmath}

\newcommand{\galex}{{\it GALEX }}
\newcommand{\planck}{{\it Planck }}
\newcommand{\ebv}{{\it E(B-V) }}

\title[]{Far-Ultraviolet diffuse emission as a tracer of Galactic extinction at the North Galactic Pole}

\author[S. Bordoloi et al.]{
Swagat Bordoloi$^{1}$,
P. Shalima$^{2}$,\thanks{E-mail: shalima.p@manipal.edu}
Rupjyoti Gogoi$^{1}$, 
Jayant Murthy$^{3}$,
Oleg Malkov$^{4}$
\\
$^{1}$Department of Physics, Tezpur University, Napaam,  784028, Assam, India\\
$^{2}$Manipal Centre for Natural Sciences, Manipal Academy of Higher Education, Manipal, India \\
$^{3}$Indian Institute of Astrophysics, Bengaluru, 560034, Karnataka, India\\
$^{4}$Institute of Astronomy, 48 Pyatnitskaya St., Moscow 119017, Russia
}

\begin{document}
\label{firstpage}
\pagerange{\pageref{firstpage}--\pageref{lastpage}}
\maketitle


\begin{abstract}

Interstellar extinction at high Galactic latitude is often traced by thermal dust emission, but the conversion from infrared surface brightness to reddening is sensitive to grain properties and temperature. We test dust-scattered far-ultraviolet (FUV) and near-ultraviolet (NUV) diffuse emission as a high-resolution extinction tracer towards the North Galactic Pole (NGP), where $E(B\!-\!V) < 0.1$~mag. Using the \textit{GALEX} diffuse maps of \cite{murthy_galex_2014}, we build FUV and NUV intensity maps at $6'$ resolution over the region $b > 70^\circ$ and compare them with the Planck $E(B\!-\!V)$ map at matched resolution. A direct pixel-to-pixel regression gives only moderate correlations ($\rho = 0.536$ for FUV, $0.243$ for NUV) owing to high per-pixel scatter from $\mathrm{H_2}$ fluorescence, line emission, and photon-counting noise; we therefore construct per-pixel \galex uncertainty maps and apply a centroid-fitting method that fits the mean intensity in bins of $E(B\!-\!V)$ after rejection of outliers in an iterative sigma-clipping method, recovering tight relations ($\rho = 0.993$ for FUV, $0.976$ for NUV). We find the best-fit relations for the FUV and NUV intensity per $6'\times6'$ pixel: $\mathrm{FUV} = 2285.3\,E(B\!-\!V) + 278.6$ and $\mathrm{NUV} = 1517.6\,E(B\!-\!V) + 547.2$ ($\mathrm{photons\,cm^{-2}\,s^{-1}\,sr^{-1}\,\mathring{A}^{-1}}$) that are inverted to produce a $601\times601$ \galex reddening map. \galex \ebv correlates strongly with the $HI$ column density from the all-sky HI4PI survey($\rho = 0.991$). We validate the maps using SDSS quasars as extragalactic extinction calibrators: quasar extinctions from the \galex and Planck maps agree with a correlation of $0.99$ and a slope of $0.986$. Comparison with the Pan-STARRS1 stellar reddening map of Schlafly et al. (2014) shows that the \galex-based values have a smaller mean difference
($0.039$~mag) than with the Planck-based values ($0.064$~mag). These results establish dust-scattered FUV diffuse emission as a viable alternative to thermal-dust extinction maps at high Galactic latitude.

\end{abstract}

\section{INTRODUCTION }
\label{sec:int}

        Interstellar extinction refers to the attenuation of starlight caused by absorption and scattering as it passes through the interstellar medium and must be corrected for in order to derive the intrinsic spectrum of any source in the sky. Accurate determination of extinction depends on the availability of nearby stars of known spectral type whose extinction can be measured as a function of wavelength \citep{fitzpatrick_correcting_1999,fitzpatrick2007analysis} and large-scale surveys used proxies such as 21 cm radiation \citep{1978bursteinandheiles} and, more recently, the infrared emission from thermally emitting dust grains \citep{schlegel_maps_1998}. The current maps are those from Planck data \citep{planck_collaboration_planck_2014} at a spatial resolution of $\sim 5'-6'$.

        However, the infrared emission is dependent on the composition and size distribution of the grains and the temperature of each component and the conversion from observed surface brightness to reddening may be variable over the sky \citep{casandjian}. \cite{boissier_galex_2015} proposed that a better tracer of extinction at higher spatial resolution would be the far-ultraviolet (FUV) observations of the Galaxy Evolution Explorer (GALEX). The dust is optically thin at high Galactic latitudes, and dust-scattered far-UV (FUV) light should serve as a better tracer of extinction at higher spatial resolution than possible from available infrared observations. 

        We have used GALEX observations to derive an extinction map around the North Galactic Pole (NGP), where the extinction is low (\ebv < 0.1 mag). The diffuse FUV here is dominated by dust-scattered starlight (the diffuse Galactic light), which is directly proportional to the reddening. Other Galactic emission from the interstellar medium — molecular-hydrogen fluorescence, two-photon continuum, and line emission, also scales with the gas column, while a large constant offset arises from the extragalactic background light and airglow \citep{akshaya_components_2019,murthy_galex_2014}. We describe the data collection and processing in Section 2 and present the data analysis in Section 3. In Section 3 we also derive the extinction of quasars from our GALEX FUV and Planck maps and examine their relation with redshift (z), and we test the reddening derived from the GALEX FUV in the following sections. We compare our results with the optical stellar observations of Schlafly et al. (2014) and the quasar extinction maps with both the GALEX and Planck maps, and finally conclude on the use of FUV emission as a tracer of extinction relative to Planck or Schlegel.

\section{Data}
\label{sec:data collection}

\begin{figure*}
    \centering
    \includegraphics[width=0.32\textwidth]{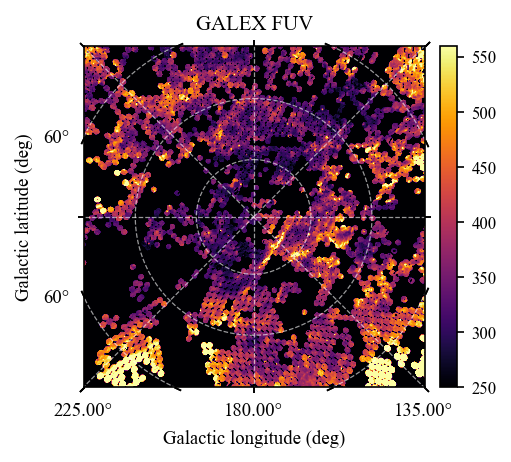}
    \hfill
    \includegraphics[width=0.32\textwidth]{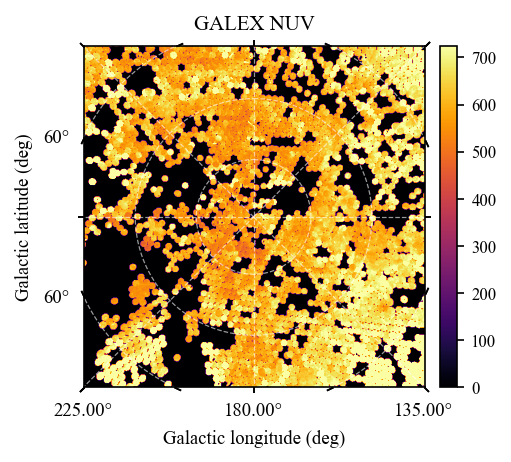}
    \hfill
    \includegraphics[width=0.32\textwidth]{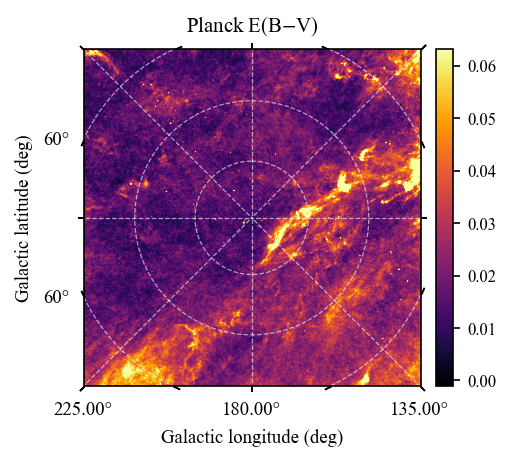}

    \caption{GALEX UV intensity maps and Planck dust reddening maps centred at the North Galactic Pole ($b = 90^\circ$). Left: GALEX Far-Ultraviolet (FUV) intensity map. Middle: GALEX Near-Ultraviolet (NUV) intensity map in native intensity units $\mathrm{photons\,cm^{-2}\,s^{-1}\,sr^{-1}\,\mathring{A}^{-1}}$. Right: Planck \ebv dust reddening map in units of magnitudes. Concentric grid lines are spaced at $10^\circ$ intervals extending outward to a Galactic latitude of $b = 60^\circ$. Angular sectors are divided into $45^\circ$ increments, going clockwise starting from $0^\circ$ at the top. Black regions in the UV maps indicate masked pixels where no valid data is available.}
\label{fig:planck-galex-2dmap}
\end{figure*}

    \subsection{Dust Emission}
        \label{dust-emission}

        The GALEX (Galaxy Evolution Explorer) mission \citep{martin_galaxy_2005,morrissey_calibration_2007} surveyed a substantial portion of the sky in two ultraviolet bands: the far-ultraviolet (FUV; $1528\mathrm{\mathring{A}}$) and the near-ultraviolet (NUV; $2360\mathrm{\mathring{A}}$). \citet{murthy_galex_2014} used these observations to generate sky maps of the diffuse cosmic radiation at $2\arcmin$ spatial resolution by masking the point sources, subtracting the foreground, and removing the airglow contribution from both bands and the zodiacal light from the NUV band. Higher-resolution ($15\arcsec$) versions of these maps are available via Zenodo%
\footnote{\url{https://doi.org/10.5281/zenodo.13337911}}.
        From these we constructed a diffuse-emission map of the NGP region ($b > 70^\circ$) at $6'$ pixel resolution, matching the resolution of the Planck observations (described below). Fig.~\ref{fig:planck-galex-2dmap} shows the resulting two-dimensional FUV and NUV intensity maps from GALEX (left and middle) in photon units, $\mathrm{photons\,cm^{-2}\,s^{-1}\,sr^{-1}\,\mathring{A}^{-1}}$.
 
        The map is centred on the North Galactic Pole ($l = 0^\circ$, $b = 90^\circ$), with concentric circles marking every $10^\circ$ in latitude and sectors divided into $45^\circ$ segments, progressing clockwise with longitude ($0^\circ < l < 360^\circ$). Darker regions indicate areas with no GALEX coverage. Both datasets reveal prominent filamentary structures in the lower-right portions of the images. These filaments were first identified in the polarization studies of \citet{hilditch1976MNRAS} and \citet{marakken1979}, and originate from dust clouds at $\sim100$--$300$~pc \citep{gontcharov2019interstellar}.
 
        We took the \ebv\ maps from the Planck mission \citep{planck_collaboration_planck_2014,planck_collaboration_planck_2016-1}. \citet{planck_collaboration_planck_2016-1} fitted these all-sky observations with the silicate--graphite--PAH dust model of \citet{draine_infrared_2007} to map the reddening over the entire sky. Planck combined the WISE and IRAS data to produce a full all-sky dust-emission map of the Milky Way at $6'$ resolution, which we resampled to match the GALEX maps (Fig.~\ref{fig:planck-galex-2dmap}, right). At the NGP the reddening is low ($\ebv \approx 0.02$--$0.05$~mag), whereas the per-pixel Planck uncertainty is $\sim5$~mmag. The uncertainty is therefore a large fraction of the reddening signal itself, so individual Planck pixels carry substantial fractional errors in the derived \ebv\ \citep{wolf2014,casandjian}. This motivates the binning and outlier-rejection approach adopted in Section~\ref{sec:data_analysis}.

\section{Data Analysis}
\label{sec:data_analysis}

\begin{figure*}
        \centering
        \includegraphics[width=6.5in]{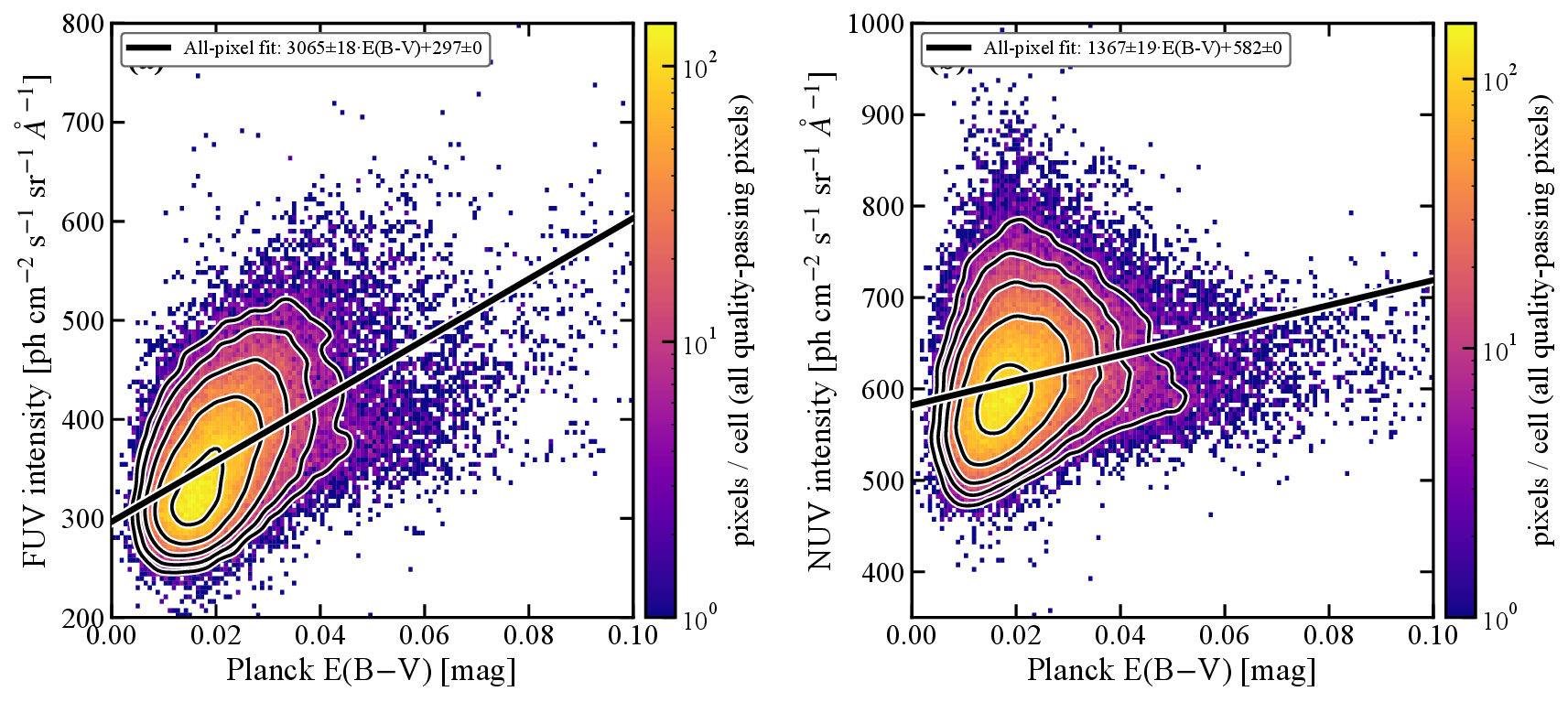}

        \caption{\galex diffuse intensity (photons\,cm$^{-2}$\,s$^{-1}$\,sr$^{-1}$\,\AA$^{-1}$) versus Planck $E(B-V)$ for the FUV (left) and NUV (right) bands, using all pixels at $b>70^\circ$. No outlier rejection has been applied. The shaded two-dimensional histogram shows the pixel density. The black line is the all-pixel least-squares fit ($\rho=0.536$ for FUV, $0.243$ for NUV); parameters are given in Table~\ref{tab:galex_planck_corr_table}.}
        \label{fig:fuv_nuv_planck_fit_colorplot}
\end{figure*}

               The Far-ultraviolet and Near-ultraviolet \galex surface brightness is correlated with the reddening from \cite{planck_collaboration_planck_2016}. We first found the correlation using all the pixels of \galex FUV and NUV map with respect to \cite{planck_collaboration_planck_2016-1}. In the first step, we only used the latitude cut; $b>70^\circ$, as shown in Fig. \ref{fig:fuv_nuv_planck_fit_colorplot}. We found a moderate correlation between FUV and Planck ($\rho=0.536$), but a lower correlation between NUV and Planck ($\rho=0.243$). The high pixel-to-pixel scatter between the GALEX and Planck maps has two distinct origins. The first is astrophysical: the GALEX bands contain additional diffuse emission that is absent from the Planck reddening map - molecular-hydrogen fluorescence, two-photon continuum, and interstellar line emission. FUV photons absorbed by molecular hydrogen ($H_2$) excite it to a higher electronic state, and the subsequent radiative de-excitation emits fluorescent photons in the FUV range ($1450$--$1750\mathrm{\mathring{A}}$; \citealt{murthy_galex_2014}). The second origin is purely statistical: the low photon-count rates of the diffuse UV background produce Poisson noise in individual pixels \cite{ananthamoorthy}. The astrophysical terms add a genuine signal that does not correlate with the Planck reddening, whereas the statistical term is random per-pixel noise; both increase the scatter and motivate the uncertainty-weighted, binned analysis described below.
                
                We constructed per-pixel uncertainty maps for both GALEX bands from the spatial scatter of the diffuse flux. On a $6'$ grid matching the intensity maps, each pixel contains several native $2'$ bins from every GALEX visit that covers it. For each visit (a single \galex pointing), we computed the variance of those bin fluxes within the pixel, and the pixel uncertainty is the square root of the mean of these per-visit variances over all visits in that pixel (Equation~\ref{eq:galex_err}).


                \begin{equation}
                    \mathrm{GALEX_{err\text{-}map}} = \sqrt{\sigma_{\rm pix}^{2}}
                   = \sqrt{\frac{1}{N_v}\sum_{v=1}^{N_v}\sigma_v^{2}},
                \label{eq:galex_err}
                \end{equation}

                \noindent where $\mathrm{GALEX_{err\text{-}map}}$ is the per-pixel uncertainty ;$\sigma_{\rm pix}^{2}$ is the per-pixel variance; $N_v$ is the number of GALEX visits covering that pixel; and $\sigma_v$ is the standard deviation of the $2\arcmin$-bin fluxes contributed by visit $v$ within the pixel, so that $\sigma_v^{2}$ is the variance of that visit. The uncertainty is given in the same photon units as the intensity maps, $\mathrm{photons\,cm^{-2}\,s^{-1}\,sr^{-1}\,\mathring{A}^{-1}}$.

\begin{figure*}
        \centering
        \includegraphics[width=6.5in]{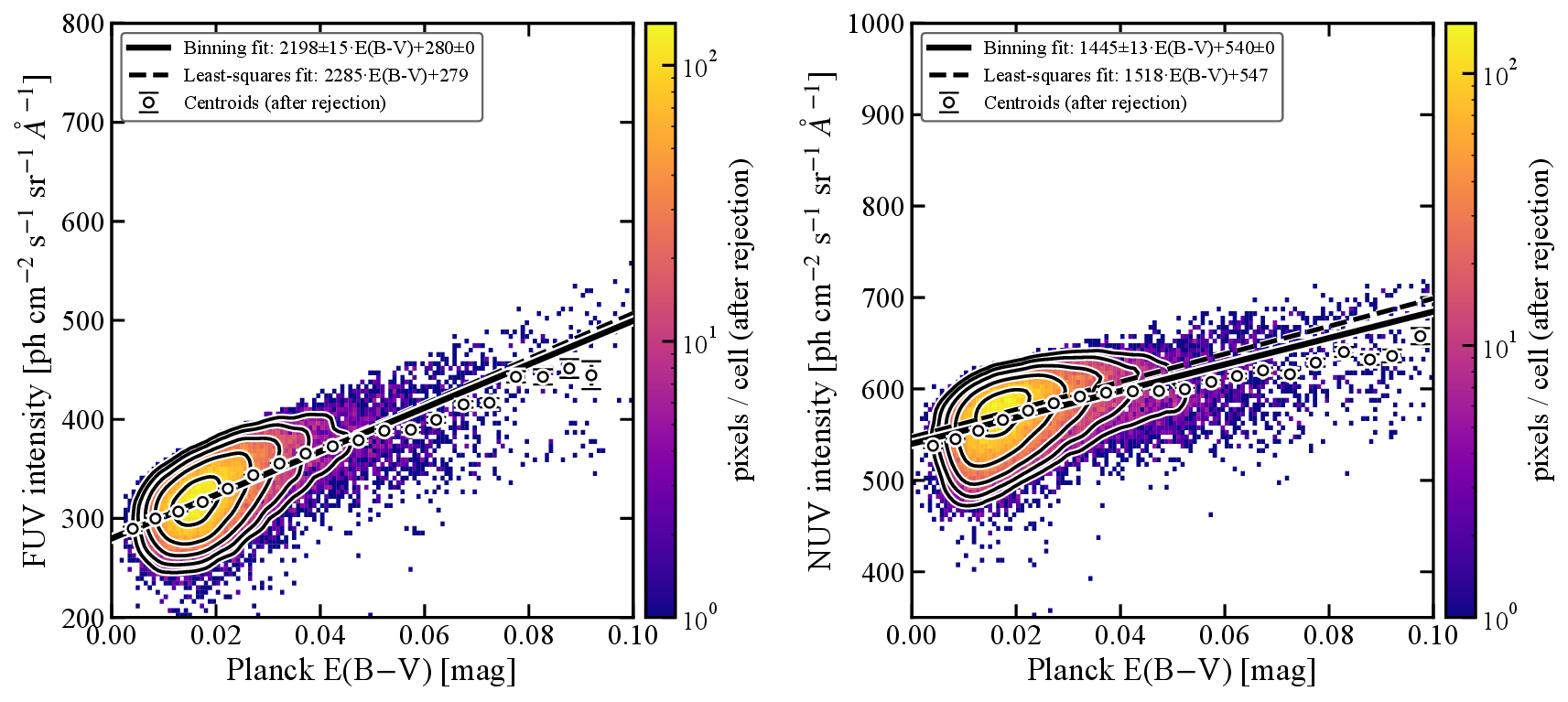}
        \caption{\galex intensity (photons\,cm$^{-2}$\,s$^{-1}$\,sr$^{-1}$\,\AA$^{-1}$) versus Planck $E(B-V)$ for the FUV (left) and NUV (right) bands, using pixels surviving iterative $\sigma$-clipping at $b>70^\circ$. The pixels shown here are the pixels that survive after the iterative clipping. White circles are the binned centroids, i.e.\ the mean \galex intensity in $0.003$\,mag bins of $E(B-V)$. The dashed line is the least-squares fit to the surviving individual pixels ($\rho=0.662$ for FUV, $0.490$ for NUV); the solid line is the fit to the binned centroids ($\rho=0.993$ and $0.976$). All parameters are listed in Table~\ref{tab:galex_planck_corr_table}.}
        \label{fig:rejection_plot_fuvnuv}
\end{figure*}

                We used the uncertainty maps of both GALEX bands, together with the Planck uncertainty map (taken directly from Planck Collaboration et al. 2016), to reject outliers as shown in Fig. \ref{fig:rejection_plot_fuvnuv}. We characterised the GALEX intensity–reddening relation in three ways. First, the all-pixel correlation: using every pixel that passed the latitude cut ($b > 70^\circ$), with no outlier rejection, we found only a moderate pixel-to-pixel correlation (FUV: $\rho$ = 0.536; NUV: $\rho$ = 0.243), owing to the large per-pixel scatter from $H_2$ fluorescence, line emission and photon-counting noise (Fig. \ref{fig:fuv_nuv_planck_fit_colorplot}). This motivates the rejection and binning steps. Second, the Least Squares method: after rejecting outliers, we performed a weighted least-squares fit to the surviving individual pixels (Fig. \ref{fig:rejection_plot_fuvnuv}), obtaining zero-reddening intercepts (offsets) of 278.6 (FUV) and 547.2 (NUV) $\mathrm{photons\,cm^{-2}\,s^{-1}\,sr^{-1}\,\mathring{A}^{-1}}$. Third, the Binning after rejection (centroid) method: we grouped the surviving pixels into equal 0.003 mag bins of \ebv, took the mean \galex intensity in each bin, and fitted these binned centroids. The parameters for all cases are listed in Table~\ref{tab:galex_planck_corr_table}. The three methods yield mutually consistent results. The correlation coefficient increases from the all-pixel value (0.536 for FUV, 0.243 for NUV) to the least-squares fit after outlier rejection (0.662, 0.490) and is highest for the binned centroids (0.993, 0.976). This progression is attributed to the successive suppression of random scatter rather than to any change in the underlying relation: the outlier rejection removes the non-dust ultraviolet excess, and the binning averages out the residual per-pixel noise, so that the correlation improves at fixed slope. The binned centroids lie almost exactly along the least-squares fit, except the NUV above 0.04 mag, where the apparent falloff is of purely statistical origin, arising from the small number of surviving pixels at high \ebv after rejection. The agreement between the centroid and per-pixel least-squares intercepts (279.8 and 278.6, respectively) confirms that the relation between \galex intensity and reddening is robust despite the substantial per-pixel scatter. We use the least-squares fit after the outlier rejection as given below.

\begin{align}
    \text{FUV} &= 2285.3 E(B-V) + 278.6 \\
    \text{NUV} &= 1517.6 E(B-V) + 547.2
    \label{eq:fuv_planck_reddening}
\end{align}

        where FUV and NUV is the \galex intensity in photon units ($\mathrm{photons ~cm^{-2}~s^{-1}~sr^{-1}}$~\AA$^{-1}$) and the reddening (\ebv) in magnitudes. We can now simply invert the equation (least squares method parameters) to derive a reddening map over the NGP.

        We prepared a $601 \times 601$ pixel \galex reddening map, covering the North Galactic Pole ($b > 70^\circ$). This map is provided at $6'$ resolution, matching that of the GALEX FUV map and Planck's reddening map, as illustrated in Fig. \ref{fig:planck-galex-2dmap}.

        We compared the extinction maps assuming a ratio of total-to-selective extinction $R_V = 3.1$. We have converted the \galex and Planck's reddening maps to their corresponding extinction maps using the standard Milky Way extinction law, using the equation:

\begin{align}
    A_V = E(B-V) \times 3.1
\end{align}

\begin{figure*}
        \centering
        \includegraphics[width=5.5in]{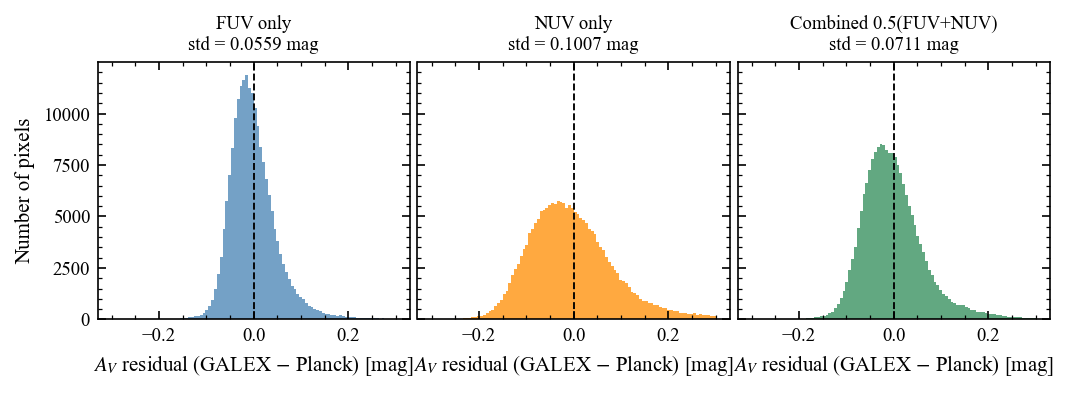}
        \caption{Residual extinction between Planck and \galex ($A_{V,Galex} - A_{V,Planck}$) in units of mag. The FUV (left), NUV (middle) and combined FUV+NUV (right) residual is compared with Planck. The NUV shows a much higher dispersion of the residual extinction values, and the combined map residual also shows a higher dispersion of the residual extinction values when compared with FUV.}
        \label{fig:residual_fuv_nuv_combined}
\end{figure*}

        We computed the residual extinction between \cite{planck_collaboration_planck_2016-1}  and \galex extinction map. For every pixel, we computed the difference in the extinction values and plotted the histogram in Fig.~\ref{fig:residual_fuv_nuv_combined}. We found that the NUV ($\sigma=0.1007$) shows a higher standard deviation than FUV only ($\sigma=0.0559$). We combined the extinction maps of \galex, giving equal weight to both the FUV and NUV bands as : 

        \begin{align}
            A_{V,FUV} = 3.1 (FUV - 278.6) / 2285.3 \\
            A_{V,NUV} = 3.1 (NUV - 547.2) / 1517.6 \\
            A_{V,combined} = 0.5 (A_{V,FUV} + A_{V,NUV})
            \label{eq:fuv_nuv_ebv_fit_parameters}
        \end{align}

        We also found the residual extinction between the combined \galex map and Planck's extinction map, and plotted the residual extinction distribution. We found that combining the FUV and NUV did not decrease the pixel-level noise, with a higher standard deviation ($\sigma=0.0711$) for the combined map as compared to FUV alone ($\sigma=0.0559$). This is because the NUV carries additional influence in light intensity from zodiacal light and airglow, and other factors.

\begin{figure}
        \centering
        \includegraphics[width=3.1in]{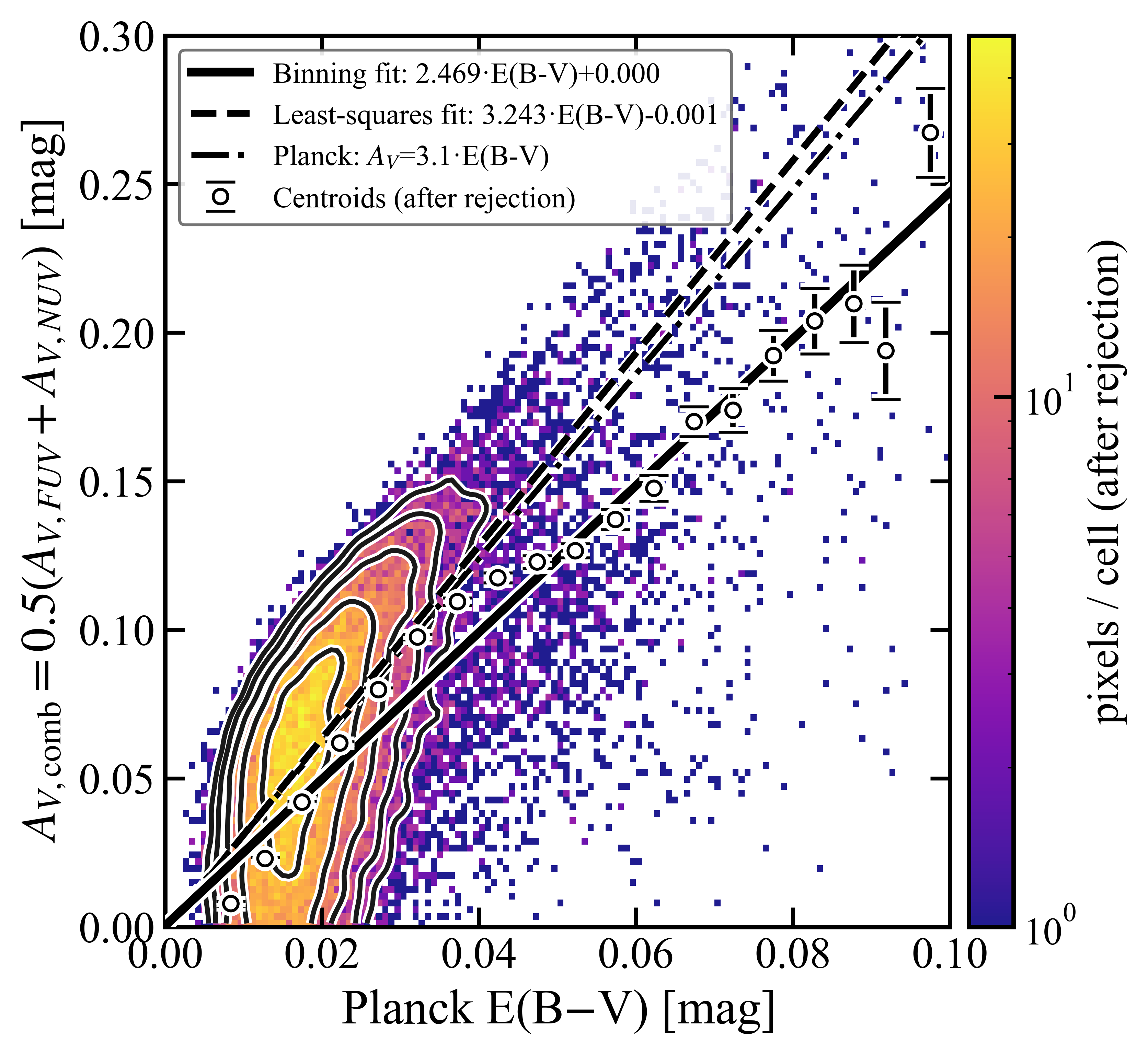}

        \caption{Combined \galex FUV$+$NUV extinction, $A_{V,\rm comb}=0.5(A_{V,\rm FUV}+A_{V,\rm NUV})$, versus Planck $E(B-V)$, for the pixels after the $\sigma$-clipping. White circles are the binned centroids, i.e.\ the mean $A_{V,\rm comb}$ in $0.003$\, mag bins of $E(B-V)$. The dashed line is the least-squares fit to the surviving pixels (slope $3.24$); the solid line is the fit to the binned centroids (slope $2.47$). The dash-dotted line is the standard Milky Way relation $A_V=3.1\,E(B-V)$ \citep{cardelli_relationship_1989}; the values consistent with least-squares fit.}
        \label{fig:fuv_nuv_combined_fit}
\end{figure}

        We fitted the combined FUV and NUV extinction map (Fig. \ref{fig:fuv_nuv_combined_fit}) versus Planck's extinction map with equal weightage in each band. After rejecting the high-uncertainty pixels, the least-squares fit yielded a slope of $3.24$, consistent with the average Milky Way value of $R_V = 3.1$ \citep{cardelli_relationship_1989}. Rejecting these outliers does indeed give us a good value close to the average Milky Way extinction law. Binning the combined FUV and NUV intensity as a function of Planck \ebv further pulls the slope from 3.24 to 2.47 because of the high pixel-level scatter in the \galex map.

        The intercept $278.6$ $\mathrm{photons~cm^{-2}~s^{-1}~sr^{-1}~\mathring{A}^{-1}}$ (FUV) and $547.2$ $\mathrm{photons~cm^{-2}~s^{-1}~sr^{-1}~\mathring{A}^{-1}}$ (NUV) represent the \galex intensity where there is no dust reddening, and mainly attributed to the integrated light from extragalactic background sources (EBL) \citep{koushan2021gama} and galactic sources \citep{alice2025}. Our value of the intercept is consistent with that of \cite{akshaya_diffuse_2018} at the NGP.
        We also compared our FUV intensity map with respect to neutral hydrogen column density ($N_{HI}$) from the HI4PI dataset \citep{HI4PI} and also validated our derived extinction map using a large sample of quasars as standard candles, as described in \cite{planck_collaboration_planck_2016-1}.

\section{Comparison with HI maps}

        We compared our FUV and Planck reddening map with the HI4PI dataset \citep{HI4PI} that maps the Galactic neutral hydrogen (HI atoms) column density, and supersedes the LAB survey. The Planck reddening map is based on the emission of the dust grains by fitting a modified blackbody, while \galex tracks the scattered light from the dust grains. $N_{HI}$ becomes an independent tracer where the HI map is derived from the $21~\mathrm{cm}$ transition of the neutral hydrogen atom. We include this comparison because $N_{HI}$, which is derived from the 21 cm line, measures the gas directly and free of any assumptions like grain composition, size distribution or temperature. So \cite{HI4PI} gives us an independent measure of the same material that \galex and Planck trace. If the FUV relation in Section 3 is real, then \galex should trace $N_{HI}$ as equally as Planck.
        
\begin{figure}
        \centering
        \includegraphics[width=3.2in]{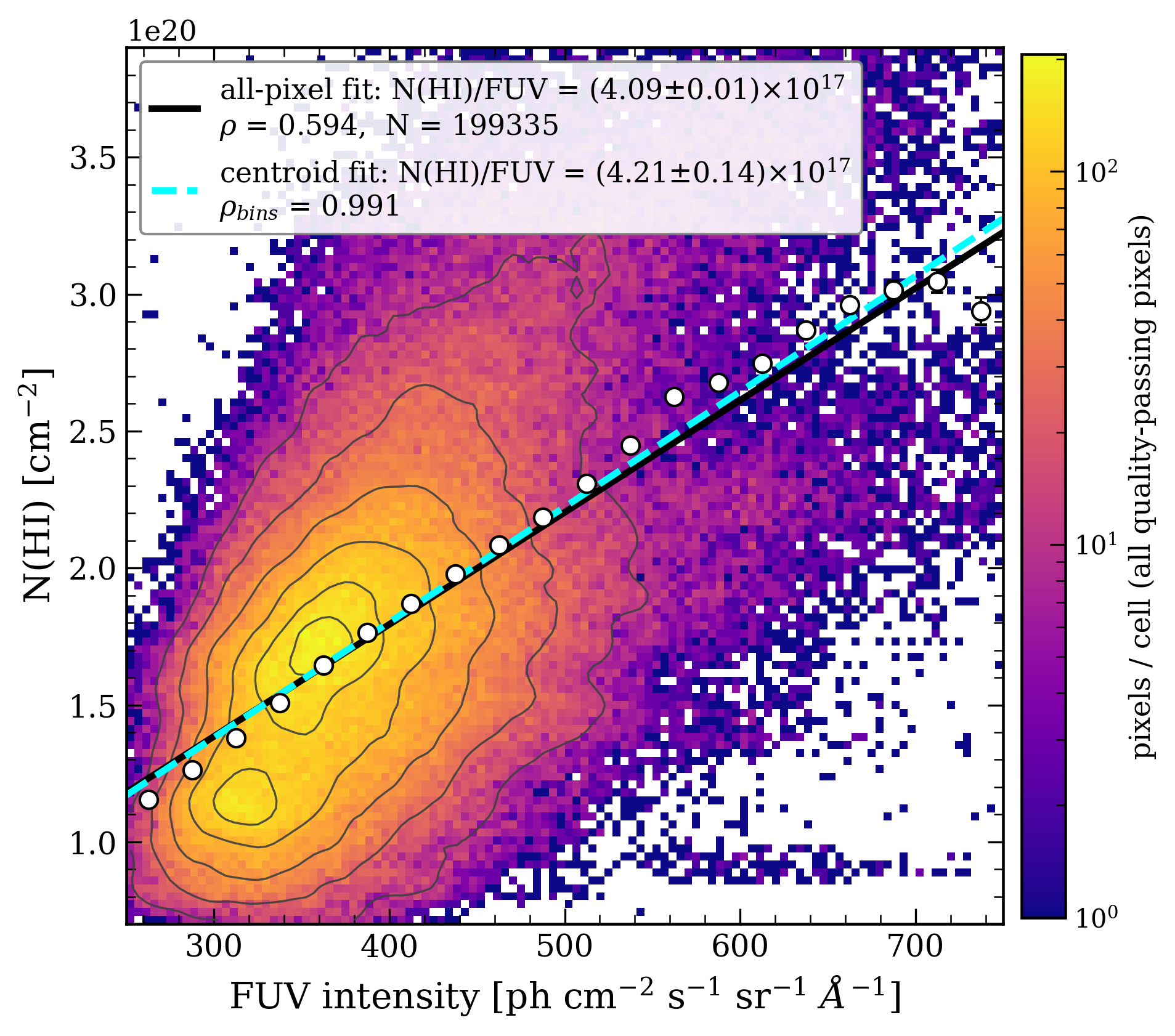}
        \caption{$HI$ column density from HI4PI versus \galex FUV intensity, for all 199\,335 quality-passing pixels at $b>70^\circ$. No $\sigma$-clipping has been applied. White circles are the binned means, i.e.\ the mean $N_{HI}$ in bins of 20 photon units of FUV intensity. The black line is the all-pixel least-squares fit ($\rho=0.594$); the cyan dashed line is the fit to the binned means ($\rho=0.991$).}
        \label{fig:hi_fuv_corr}
\end{figure}

\begin{figure}
        \centering
        \includegraphics[width=3.2in]{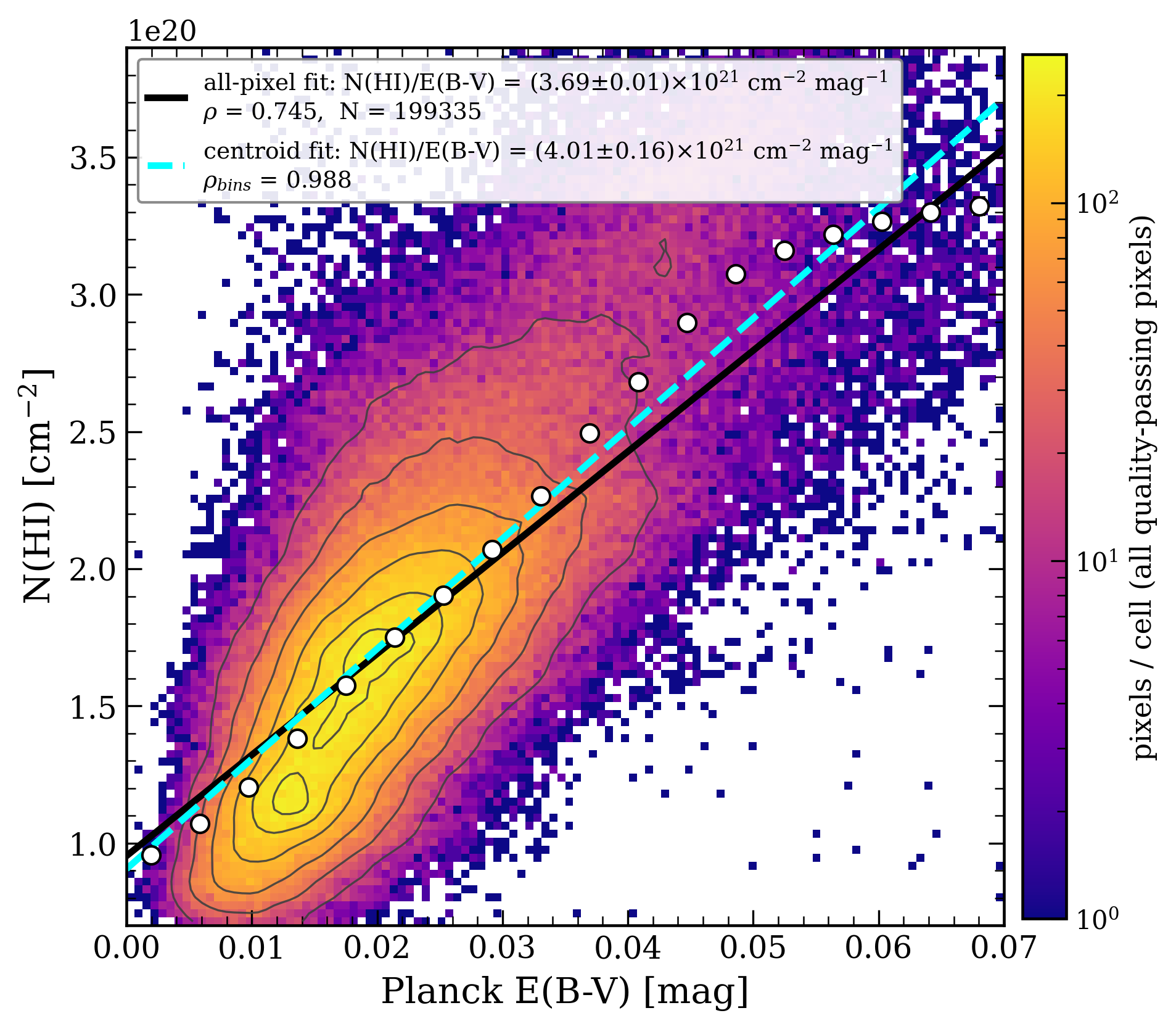}
        \caption{$HI$ column density from HI4PI versus Planck $E(B-V)$, with the binned means taken in $0.003$\,mag bins of $E(B-V)$ for pixels at $b>70^\circ$. No $\sigma$-clipping has been applied. The black line is the all-pixel least-squares fit ($\rho=0.745$) and the cyan dashed line is the binned-mean fit ($\rho_{\rm bins}=0.988$).}
        \label{fig:hi_planck_corr}
\end{figure}

        Unlike the GALEX–Planck comparison in Section 3, no sigma-clipping rejection was applied here: both fits below use the full sample of pixels ($N = 199,335, ~b>70^\circ$). We used the centroid-fitting algorithm between the $N_{HI}$ map from the HI4PI survey and, separately, the FUV intensity and Planck reddening. The ratios $N_{HI}/\mathrm{FUV}$ and $N_{HI}/E(B-V)$ are given in Table~\ref{tab:hi-galex_planck_corr_table}. We plotted the mean $HI$ column density binned at equal intervals of 20 photon units of FUV intensity (Fig.~\ref{fig:hi_fuv_corr}), and the mean column density in $0.003$\ mag bins of Planck reddening    (Fig.~\ref{fig:hi_planck_corr}), finding a strong linear correlation for both \galex and Planck \citep{planck_collaboration_planck_2016-1} with respect to the neutral hydrogen column density. We find $N_{HI}/E(B-V) = (4.01\pm0.16)\times10^{21}$\,atoms\,cm$^{-2}$\,mag$^{-1}$, slightly lower than the $5.8\times10^{21}$\,atoms\,cm$^{-2}$\,mag$^{-1}$ obtained by \citet{bohlin1978}, who sampled OB stars in the denser Galactic plane. For the FUV we find $N_{HI}/\mathrm{FUV} = (4.21\pm0.14)\times10^{17}$\,atoms\,cm$^{-2}$\,(photons\,cm$^{-2}$\,s$^{-1}$\,sr$^{-1}$\,\AA$^{-1}$)$^{-1}$.

        Additionally, we found the correlation between the \galex intensity and Planck's reddening with respect to neutral hydrogen $N_{HI}$ using pixel-to-pixel cross-correlation, and compared with the centroid fitting algorithm. We found a much lower correlation of both the \cite{planck_collaboration_planck_2016-1} and \galex map with respect to hydrogen atom column density ($N_{HI}$): $\rho=0.594$ for GALEX and $\rho=0.745$ for Planck. This is solely attributed to the noise of the individual pixels of \galex and Planck and genuine sightline variations in the datasets. Once the pixels are binned, both the GALEX and Planck correlations with $N_{HI}$ recover to $\rho=0.991$ and $\rho=0.988$, respectively, confirming that the scatter is dominated by per-pixel noise.

   \begin{table*}
    \centering
    \begin{tabular}{l|c|c|c}
    \hline
    $A_V$ maps & Slope & Intercept & Correlation ($\rho$)\\ 
    \hline
    & $\mathrm{photons~cm^{-2}~sr^{-1}~s^{-1}~\mathring{A}^{-1}~mag^{-1}}$ 
    & $\mathrm{photons~cm^{-2}~sr^{-1}~s^{-1}~\mathring{A}^{-1}}$ & \\
    \hline
    FUV vs. Planck (All pixel)    & $3064.5\pm18.4$  & $296.7\pm0.4$      & 0.536 \\
    FUV vs. Planck (Binning after rejection) & $2198\pm15$ & $279.8\pm3$  & 0.993 \\
    FUV vs. Planck (Least Square) & $2285.3$ & $278.6$  & 0.662 \\
    \hline
    NUV vs. Planck (All pixel)    & $1367.4\pm19.2$  & $582\pm0.5$      & 0.243 \\
    NUV vs. Planck (Binning after rejection) & $1445\pm13$ & $540$  & 0.976 \\
    NUV vs. Planck (Least Square) & $1517.6$ & $547.2$  & 0.490 \\
    \hline
    \end{tabular}
    \caption{Fit parameters of the extinction relation between GALEX and Planck. Slope units are photons cm$^{-2}$ sr$^{-1}$ s$^{-1}$ \AA$^{-1}$ mag$^{-1}$ and intercept units are photons cm$^{-2}$ sr$^{-1}$ s$^{-1}$ \AA$^{-1}$. We provide three fits: the all-pixel correlation (no rejection), the least-squares fit to the individual surviving pixels after rejection, and the centroid (binning-after-rejection) fit to the binned means. }
    \label{tab:galex_planck_corr_table}
\end{table*}

   \section{Estimating \texorpdfstring{$A_{V,QSO}$}{AV,QSO} using Sloan Digital Sky Survey Quasar Catalog}

        We followed the methodology described in \cite{planck_collaboration_planck_2016-1} to derive the quasar extinction $A_{V,QSO_{Planck}}$ at the North Galactic Pole, utilising quasars from the Sloan Digital Sky Survey as standard candles. The Planck collaboration renormalised extinction values from the \cite{draine_infrared_2007} model to match quasar-based extinction estimates.

        We used the SDSS Data Release 7 quasar catalog from \cite{schneider2010sloan}, comprising approximately 105,783 quasars, complemented by 166,583 spectroscopically confirmed quasars from \cite{paris2014}. These catalogs provide photometric measurements across the five \textit{ugriz} bands, as well as precise redshift estimates. As emphasised by \cite{planck_collaboration_planck_2016-1}, quasars serve as effective calibrators owing to several advantages, foremost among them their extragalactic origin and high redshifts. 

\begin{table*}
    \centering
    \begin{tabular}{l|c|c}
    \hline
    $A_V$ maps & Ratio (Units) & Correlation ($\rho$)\\ 
    
    \hline
    $N_{\rm HI}/\rm{FUV}$ (All pixel) & $(4.09 \pm 0.01)\times 10^{17}$ 
    atoms cm$^{-2}$ (photons cm$^{-2}$ s$^{-1}$ sr$^{-1}$ \AA$^{-1}$)$^{-1}$ & 0.594\\
    
    $N_{\rm HI}/\rm{FUV}$ (Binning)& $(4.21 \pm 0.14)\times 10^{17}$ 
    atoms cm$^{-2}$ (photons cm$^{-2}$ s$^{-1}$ sr$^{-1}$ \AA$^{-1}$)$^{-1}$ & 0.991\\
    \hline
    $N_{\rm HI}/E(B-V)$ (All pixel)  & $(3.69 \pm 0.01)\times 10^{21}$ 
    atoms cm$^{-2}$ mag$^{-1}$ & 0.745\\

    $N_{\rm HI}/E(B-V)$ (Binning)  & $(4.01 \pm 0.16)\times 10^{21}$ 
    atoms cm$^{-2}$ mag$^{-1}$ & 0.988\\
    \hline
    \end{tabular}
    \caption{All-pixels fit and centroid binning fit parameters of N(HI) column density from HI4PI 
    with respect to GALEX FUV intensity and Planck $E(B-V)$ reddening. 
    The N$_{\rm HI}$/FUV ratio is in units of 
    atoms cm$^{-2}$ (photons cm$^{-2}$ s$^{-1}$ sr$^{-1}$ \AA$^{-1}$)$^{-1}$ 
    and the N$_{\rm HI}$/$E(B-V)$ ratio is in units of 
    atoms cm$^{-2}$ mag$^{-1}$.}
    \label{tab:hi-galex_planck_corr_table}
\end{table*}

        Leveraging the high-precision photometry from SDSS, we performed a parallel comparison of quasar extinctions using Planck ($A_{V,\mathrm{Planck}}$) and \galex ($A_{V,\galex}$) 2D extinction maps. For this, we first determined the intrinsic colours of unreddened quasars using the reddening maps. Assuming the dust extinction curve from \cite{fitzpatrick_correcting_1999}, we averaged the extinction values across the colour groups into a single quasar extinction value.

\begin{figure}
    \centering
    \includegraphics[width=3in]{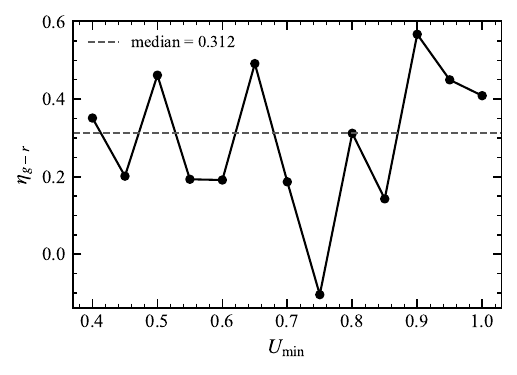}
    \caption{Slope $\eta_{g-r}$ of the linear fit between observed magnitude ($m_X - m_Y$) as a function of $U_{min}$ applied to the \galex reddening map. The dashed line marks the median across all $U_{min}$ values ($\eta_{g-r}=0.312$).}
    \label{fig:eta_umin}
\end{figure}

\begin{figure*}
    \centering
    \includegraphics[width=5.5in]{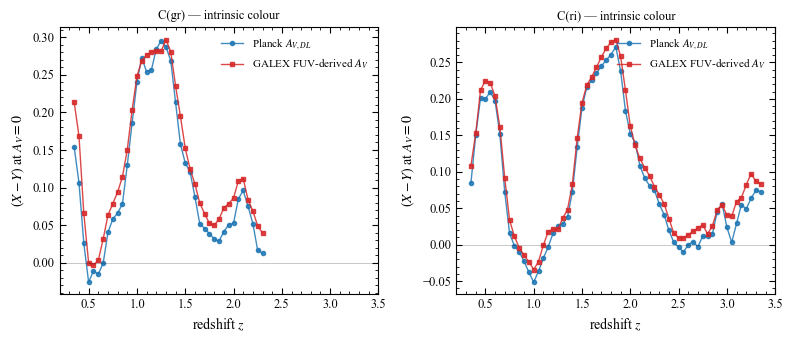}

\caption{Intrinsic QSO colours $C_{g,r}(z)$(left) and $C_{r,i}(z)$(right) as a function of redshift, obtained from zero extinction intercept ($A_{V,GALEX}=0$) of the linear fit between observed QSO colours and derived extinction from the maps. Blue points show Planck-derived intrinsic colours of the QSO, while red points show \ \galex-derived intrinsic colours. The two maps agree closely across the redshift range.}
\label{fig:intrinsic_color_vs_redshift}
\end{figure*}

    \subsection{Quasar catalog}


        The Sloan Digital Sky Survey \citep{gunn2006} operates using a 2.5-m modified Ritchey-Chrétien altitude-azimuth telescope mounted at Apache Point Observatory. It conducts observations in five optical bands ($ugriz$), featuring effective wavelengths of 359.49 nm ($u$), 464.03 nm ($g$), 612.23 nm ($r$), 743.95 nm ($i$), and 889.70 nm ($z$) \citep{fukugita}. The SDSS Data Release 7 quasar catalog, which includes 105,783 spectroscopically confirmed quasars, was released by \cite{schneider2010sloan} and later expanded by \cite{paris2014} with an additional 166,583 quasars.

        To remove potential biases from $Ly\alpha$ emission in our samples, we implemented a series of selection criteria. We adopted the redshift thresholds specified in \cite{planck_collaboration_planck_2016-1}: $z < 1.64$ ($u$), $z < 2.31$ ($g$), $z < 3.55$ ($r$), $z < 4.62$ ($i$), and $z < 5.69$ ($z$). Additionally, we constructed a subsample over the redshift interval $0.35 < z < 3.35$ to determine intrinsic quasar colours. \cite{planck_collaboration_planck_2016-1} accounted for the variations in the interstellar medium thermal properties based on the starlight heating intensity ($U_{min}$), which Planck used to model the dust thermal emission, following the previous models of \cite{draine_infrared_2007}. Most of the dust is heated by an interstellar radiation field $U_{min}$, and it traces the typical heating of the cold diffuse dust along each line of sight. Finally, we restricted the datasets to Galactic latitudes $b > 70^\circ$ for analysis centred on the North Galactic Pole.

    \subsection{Estimating the intrinsic colour}

        The intrinsic colour of a quasar is strongly redshift-dependent. The shifting of the quasar spectrum due to the redshift ($z$) causes the continuum and emission lines to move through the fixed SDSS passbands. 
        
        We binned the quasars in the redshift interval $[z-0.2,z+0.2]$ in the entire catalog, and fitted the colours of quasars $(m_X-m_Y)$ as a function of extinction from the \galex map ($A_{V,\galex}$) in each respective bin. Bins containing fewer than 10 quasars were excluded to get a good estimate of the fit. For each bin, we fitted the observed colour ($m_X - m_Y$) as a linear function of extinction $A_{V,\galex}$ as:

\begin{equation}
(m_X-m_Y) = C_{X,Y}(z,U_{min}) +  \eta_{X,Y}(z,U_{min})A_{V,\galex}
\label{binning_avgalex}
\end{equation}

            where $C_{X,Y}$ is the intrinsic colour, $(m_X-m_Y)$ is the observed colour of the quasar. $A_{V,\galex}$ is the extinction value along the line of sight extracted from the \galex \ebv map, and converted using the standard Milky Way extinction law with $R_V=3.1$. The intercept represents the intrinsic colour $C_{X,Y}(z)$, when the source is un-obscured.

            We checked the effect of $U_{min}$ parameter on the slope ($\eta_{X,Y}$). We grouped the QSOs in bins of $U_{min}$ ranging from 0.4 to 1.0, and fit the observed colour with respect to the extinction from \galex. We noted down the slope in each $U_{min}$ interval. Fig. \ref{fig:eta_umin} shows a large scatter or variation of $\eta_{X,Y}$ with respect to $U_{min}$. Unlike the variation found by \cite{planck_collaboration_planck_2016-1} over the full sky, we found no significant trend of $\eta$ with $U_{min}$, with values scattering around a median of 0.312 over the range $0.4 \le U_{min} \le 1.0$. This shows that the heating intensity over the North Galactic Pole is uniform and does not depend on $U_{min}$. Therefore, we grouped all the $U_{min}$ values together to fit the observed colour $(m_X - m_Y)$ with the extinction from \galex as:

\begin{equation}
(m_X-m_Y) = C_{X,Y}(z) +  \eta_{X,Y}(z)A_{V,\galex}
\label{new_binning_avgalex}
\end{equation}

        After determining $C_{X,Y}$ for all band pairs across the entire redshift range, we obtained a single estimate of the intrinsic colours of the SDSS band pairs, $C_{X,Y}$. We plotted the intrinsic colours of the $g,r,i,z$-bands as a function of redshift in Fig.~\ref{fig:intrinsic_color_vs_redshift}. We found that the intrinsic colours of the QSOs using the Planck and \galex as references obtained very similar results to Planck.

            We computed the colour excess, or the reddening in band pair $E_{X,Y}$ as:
\begin{equation}
    E_{X,Y} = (m_X-m_Y) - C_{X,Y}(z)
\label{exy_derivation}
\end{equation}
            where $(m_X-m_Y)$ is the observed colours of the quasar.

\begin{figure}
    \centering
    \includegraphics[width=2.9in]{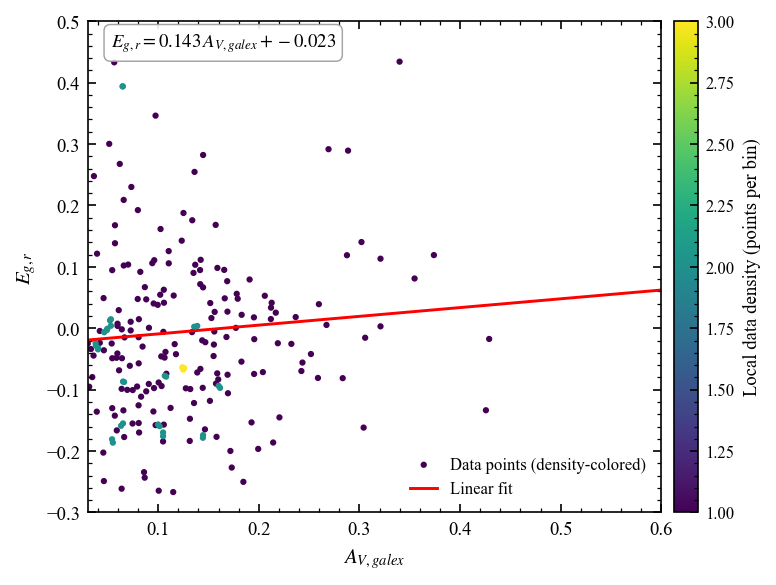}
    \caption{Linear regression fit overlaid on a density-coloured scatter plot of $E_{g,r}$ versus $A_{V,\galex}$. The best-fit line follows the equation $E_{g,r} = 0.143 A_{V,\mathrm{galex}} - 0.023$.}
    \label{fig:egr_vs_avgalex}
\end{figure}

        Fig. \ref{fig:egr_vs_avgalex} shows the colour excess $E(g-r)$ vs extinction from \galex in low-extinction regime. The map shows a large scatter, showing variation in the quasar spectra that dominates the dust signal in low-extinction regions. The large scatter is mainly attributed to the variation in the intrinsic colour of the quasar at fixed redshifts (in the low-extinction regime), set by variation in properties like the intrinsic dust reddening within the quasars. Caution is therefore required when applying the map to individual sightlines, although it remains reliable for ensemble-averaged measurements. We used the mean intrinsic colour $C_{X,Y}$ that we derived using a large sample of QSOs over a redshift range, and then derived the colour excess of the quasars using Equation \ref{exy_derivation}.

    \subsection{Estimating $A_{V,QSO_{\galex}}$}

        To derive the quasar extinction $A_{V,{QSO_{\galex}}}$ in the standard Johnson $V$ band, we applied the conversion factor $\delta_{X,Y}$ as shown by \citet{planck_collaboration_planck_2016-1}. Extinctions were computed for quasars across all SDSS band pairs and subsequently averaged to obtain a single estimate of $A_{V,{QSO_{\galex}}}$.

        To calculate the conversion factor, we computed the relative extinction in the SDSS $ugriz$ bands per unit extinction in the Johnson $V$ band ($\lambda = 540$ nm), denoted $A_X / A_V$. This calculation uses the composite quasar spectrum from \cite{vanden2001composite} and the extinction curve of \cite{fitzpatrick_correcting_1999}, parameterized with $R_V = 3.1$.

\begin{table}
\centering

\begin{tabular}{cccccc}
\hline
$z$ & $A_u/A_V$ & $A_g/A_V$ & $A_r/A_V$ & $A_i/A_V$ & $A_z/A_V$ \\
\hline
0.5 & 1.558 & 1.213 & 0.832 & 0.619 & 0.451 \\
1.0 & 1.559 & 1.203 & 0.836 & 0.621 & 0.456 \\
1.5 & 1.556 & 1.211 & 0.831 & 0.622 & 0.462 \\
2.0 & 1.541 & 1.207 & 0.838 & 0.618 & 0.463 \\
2.5 & 1.549 & 1.211 & 0.831 & 0.618 & 0.458 \\
\hline
\end{tabular}
\caption{Relative extinction $\delta_X(z,\mathrm{R_V}=3.1) = A_X/A_V$ in the SDSS $u,g,r,i,z$ bands, per unit extinction in the Johnson $V$ band, from the \citet{fitzpatrick_correcting_1999} extinction curve ($\mathrm{R_V = 3.1}$) applied to the composite QSO spectrum of \citet{vanden2001composite}, at representative redshifts spanning the fitted range.}

\label{tab:axav_vs_redshift}
\end{table}

\begin{figure}
    \centering
    \includegraphics[width=2.5in]{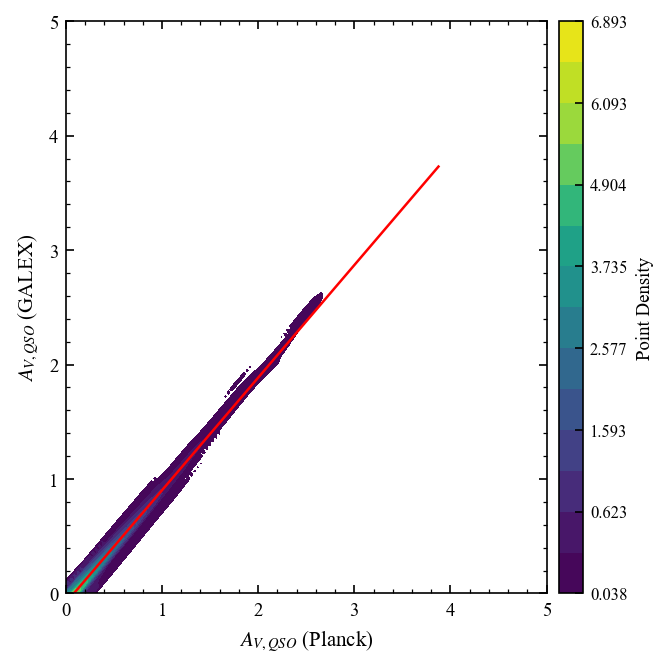}
    \caption{The straight line fit between QSO extinctions measured using \galex and Planck. The mean fit is shown by the red coloured line.}
    \label{fig:avqso_avqsogalex_fit}
\end{figure}      

            We calculated the ratio $A_X/A_V$ as:

            \begin{equation}
                \delta_X(z,R_V=3.1) = \frac{A_X}{A_V}
            \end{equation}

            The $A_X/A_V$ has been tabulated for all the filters of SDSS $ugriz$ bands in Table \ref{tab:axav_vs_redshift}. Using the ratios, we calculated the extinction of the quasar $A_{V,QSO_{galex}}$ as:
            
            \begin{equation}
             A_{V,QSO_{galex}} = \delta_{X,Y}~\times~E_{X,Y}   
            \end{equation}

            where,

            \begin{equation*}
             \delta_{X,Y} = \frac{1}{\delta_X(z,R_V=3.1) - \delta_Y(z,R_V=3.1)}   
            \end{equation*}

            Finally, we derived a single quasar extinction estimate by averaging the values across all SDSS band pairs. The same procedure was followed to obtain the quasar extinction values ($A_{V,{QSO_{\planck}}}$) using the Planck map ($A_{V,{Planck}}$).

\begin{figure*}
    \centering
    \includegraphics[width=6.5in]{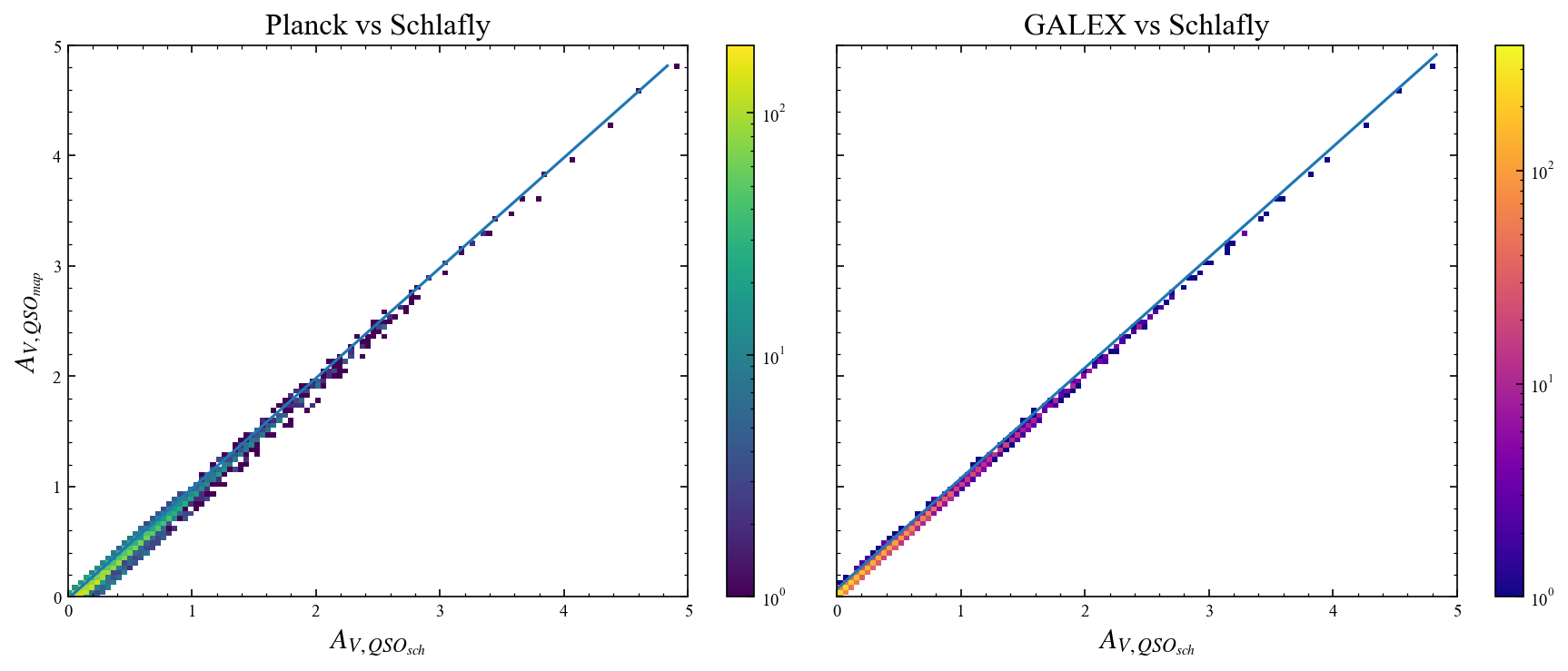}
    \caption{QSO extinction measured using \galex in comparison with stellar observations of Pan-STARRS1 by \protect\cite{schlafly2014}(right). Side by side is plotted with Planck (left) and Schlafly et al. The mean fit of the extinction values of the quasars is shown by the blue line.}
    \label{fig:avqso_sch}
\end{figure*}

        The relation between the Planck and \galex QSO extinction values exhibits a strong correlation coefficient of $0.99$ and a slope of $0.986$. The fit parameters are detailed in Table~\ref{tab:extinction correlation planck and galex}, and the fit is shown in Fig.~\ref{fig:avqso_avqsogalex_fit}.

    \begin{table}
        \centering
        \begin{tabular}{c|c|c|c}
        \hline
            Quasar extinction fit & Slope & Intercept & Correlation ($\rho$) \\
            \hline
              &&&\\$A_{V,QSO_{GALEX}}$ vs $A_{V,QSO_{Planck}}$ & 0.986 & -0.088 & 0.99\\
             &&&\\
             \hline
        \end{tabular}
        \caption{The best-fit parameters of QSO extinctions between \galex and \planck.}
        \label{tab:extinction correlation planck and galex}
    \end{table}

    \section{Validation of quasar extinctions using reddening maps from stellar observations}

            We compared the quasar extinction values derived in the previous sections with those obtained from the reddening map of \cite{schlafly2014}. This map was constructed using Pan-STARRS1 stellar photometry for the full northern Galactic region ($b > -30^\circ$) at angular resolutions of $7'-14'$ extending to 4.5 kpc, based on distance and reddening estimates for over 500 million stars. Following the same procedure as outlined in the previous sections, we derived the quasar extinction $A_{V,{QSO_{sch}}}$ using the \cite{schlafly2014} reddening map for latitudes $b > 70^\circ$.

            Fig. \ref{fig:avqso_sch} compares the quasar extinctions derived from the \galex and Planck maps with stellar extinctions from the map of \cite{schlafly2014} in the region around the North Galactic Pole ($b>70^\circ$). The results are summarised in Table \ref{tab:quasar_extinction_schlafly}. Excellent agreement was found between the quasar extinction estimates $A_{V,{QSO_{sch}}}$ from the Schlafly map and those from the two maps ($A_{V,{QSO_{\galex}}}, A_{V,{QSO_{Planck}}}$). The mean difference between the \galex and Schlafly quasar extinctions is lower ($A_{V,{QSO_{\galex}}} - A_{V,{QSO_{sch}}}=0.0390$ mag), with a standard deviation of $0.0185$ mag, compared to the Planck and Schlafly ($A_{V,{QSO_{Planck}}} - A_{V,{QSO_{sch}}}=0.0639$ mag; mean standard deviation of $0.0341$ mag).

            This analysis validates the use of far-ultraviolet diffuse emission as a tracer of extinction at high Galactic latitudes ($b>70^\circ$), where the \galex data exhibit a comparable, but lower standard deviation and mean difference as compared to \cite{planck_collaboration_planck_2016-1}.

\begin{table}
    \centering
    \resizebox{\columnwidth}{!}{%
    \begin{tabular}{c|c|c|c|c}
    \hline
     $A_{V,QSO_{X}}$ vs $ A_{V,QSO_{Y}}$ & Slope & Intercept & Std Dev & $A_{V,QSO_{X}} - A_{V,QSO_{Y}}$\\
    & &  &   & \\
    \hline
        GALEX vs. Schlafly & 0.99 & 0.03 & 0.0185 & 0.0390\\
        Planck vs. Schlafly & 0.98 & -0.07 & 0.0341 & 0.0639\\
    \hline         
    \end{tabular}
    }
    \caption{The best-fit parameters of quasar extinctions of Schlafly versus Planck and GALEX. The last column shows the mean difference between the extinction values of quasars between the two maps, where $X$ and $Y$ are the extinction maps, respectively.}
    \label{tab:quasar_extinction_schlafly}
\end{table}

\section{Conclusion}

We have constructed a far-ultraviolet (FUV) extinction map of the North Galactic Pole region ($b > 70^\circ$), where the reddening is low (\ebv $< 0.1$ mag), and assessed dust-scattered FUV diffuse emission as a tracer of Galactic extinction. A direct pixel-to-pixel regression yielded only moderate correlations
($\rho = 0.536$ for FUV and $\rho = 0.243$ for NUV). This scatter reflects the additional per-pixel contributions from molecular-hydrogen fluorescence, interstellar line emission, and photon-counting noise rather than a weakness of the underlying relation. To recover the intrinsic relation, we constructed per-pixel GALEX uncertainty maps. We applied a centroid-fitting method to estimate the mean intensity in \ebv\ bins after iteratively sigma-clipping outliers. This increased the correlations to $\rho = 0.993$ (FUV) and $\rho = 0.976$ (NUV) and gave the best-fit relations $\mathrm{FUV} = 2285.3\,\ebv + 278.6$ and
$\mathrm{NUV} = 1517.6\,\ebv + 547.2$ (photon units). The fitted offsets are consistent with the non-reddening-correlated background (extragalactic background light and airglow) reported in previous works \citep{akshaya_diffuse_2018,akshaya_components_2019}. Inverting these relations, we produced a $601 \times 601$ GALEX-based reddening map of the NGP.
 
We validated the maps in three independent ways. First, both the GALEX and Planck maps correlate strongly with the \ion{H}{i} column density from the HI4PI survey \citep{HI4PI} ($\rho = 0.991$ and $0.988$, respectively), confirming that the FUV diffuse emission traces the gas and dust column. Second, using SDSS quasars as extragalactic extinction calibrators, the quasar extinctions derived from the GALEX and Planck maps agree closely (correlation $0.99$, slope $0.986$). Third, a comparison with the Pan-STARRS1 stellar reddening map of \citet{schlafly2014} shows that the GALEX-based reddening has a smaller mean difference ($0.039$~mag, $\sigma = 0.019$) than the Planck-based reddening ($0.064$~mag, $\sigma = 0.034$), indicating that the FUV-derived map is at least as accurate as the thermal-dust map at high Galactic latitude.

    To show that the \galex reddening map provides comparable extinction estimates to the Planck map, we used the quasar catalog from \cite{schneider2010sloan, paris2014}. As the quasars are extragalactic and distributed across the sky, they serve as ideal extinction calibrators. Following the procedure outlined in \cite{planck_collaboration_planck_2016-1}, we estimated the quasar intrinsic colours, which depend strongly on redshift ($z$). Fig. \ref{fig:intrinsic_color_vs_redshift} shows that the \galex map shows good agreement with Planck in computing the band-pair intrinsic colours ($C_{X,Y}$) relative to redshift. We computed $C_{X,Y}$ over the redshift range, avoiding Ly$\alpha$ leakage in the SDSS $ugriz$ filters. Reddening values $E_{X,Y}$ for all SDSS band pairs ($X,Y$) were then derived by subtracting intrinsic colours from observed colours. We find a large scatter of the QSO intrinsic colour while comparing with the map's extinction values at the position. The intrinsic colour of the QSO at fixed redshift varies in the continuum and the spectra, leading to the large scatter in Fig. \ref{fig:egr_vs_avgalex}, but the strong agreement between the QSO extinctions derived from \galex and Planck says that the \galex map can be used when averaged over an ensemble of quasars or stellar sources.

    Taken together, these results establish dust-scattered FUV diffuse emission as a viable alternative to thermal-dust extinction maps in the optically thin, low-extinction regime at high Galactic latitude. Because the scattered FUV light is directly proportional to the reddening and is independent of grain temperature and emission, the way the Planck and Schlegel's map was prepared, it offers a physically distinct and complementary method to map Galactic extinction.

    \subsection{Future work}
        
        A potential advantage of far-ultraviolet diffuse emission as an extinction tracer is its finer native pixel scale compared to Planck's data. Planck achieves a maximum resolution of approximately $6'$, whereas \galex \citep{morrissey_calibration_2007,jmurthy2025} provides far-ultraviolet diffuse emission maps at a resolution of $15''$. As part of future efforts, we plan to extend the analysis to regions below $b = 70^\circ$, testing whether the FUV--Planck relation derived here for the North Galactic Pole holds at other, lower Galactic latitudes. This would allow dereddening corrections to be validated over a larger sky area, and facilitate analysis of dust properties using an independent tracer relative to Planck. The \galex and Planck reddening maps of the North Galactic Pole produced in this work are publicly available at \url{https://github.com/swagatastro98/NGP_resources.}

\section{ACKNOWLEDGEMENT}

     This research has made use of data from the Galaxy Evolution Explorer (\galex) \cite{morrissey_calibration_2007} and the diffuse \galex data from \cite{murthy2016m}. We also obtained the Planck Satellite \cite{planck_collaboration_planck_2014, planck_collaboration_planck_2016-1} data from \footnote{\url{https://pla.esac.esa.int/}}. We also thank the Sloan Digital Sky Survey (SDSS) \cite{sdss_almeida} collaboration for providing their valuable data. We use the GNU Data Language (GDL) \cite{coulais2025gdl} and Python v3.10.16 for our analysis.

\section*{Data Availability}
    The diffuse FUV data derived from GALEX observations are available in the repository and can be accessed via the Zenodo url: \url{https://doi.org/10.5281/zenodo.13337911}. Planck mission data is available from their respective mission archives at \url{https://pla.esac.esa.int/pla/}.

\bibliographystyle{mnras}
\bibliography{references}

\label{lastpage}
\end{document}